# Star formation near the Sun is driven by expansion of the Local Bubble

Catherine Zucker[1,2,*], Alyssa A. Goodman[1], João Alves[3], Shmuel Bialy[1,4], Michael Foley[1], Joshua S. Speagle[6,7,8], Josefa Groβschedl[3], Douglas P. Finkbeiner[1,5], Andreas Burkert[9,10], Diana Khimey[1], Cameren Swiggum[3,11]

[1] Center for Astrophysics | Harvard & Smithsonian, 60 Garden St. Cambridge, MA 02138, USA
[2] Space Telescope Science Institute, 3700 San Martin Drive, Baltimore, MD 21218, USA
[3] University of Vienna, Department of Astrophysics, Türkenschanzstraße 17, 1180 Wien, Austria
[4] Department of Astronomy, University of Maryland, College Park, MD 20742, USA;
[5] Harvard University Department of Physics, 17 Oxford Street, Cambridge, MA 02138, USA
[6] Department of Statistical Sciences, University of Toronto, Toronto, ON M5S 3G3, Canada
[7] David A. Dunlap Department of Astronomy & Astrophysics, University of Toronto, Toronto, ON M5S 3H4, Canada
[8] Dunlap Institute for Astronomy & Astrophysics, University of Toronto, Toronto, ON M5S 3H4, Canada
[9] University Observatory Munich (USM), Scheinerstrasse 1, 81679 Munich, Germany
[10] Max-Planck-Institut für extraterrestrische Physik (MPE), Giessenbachstr. 1, 85748 Garching, Germany
[11] Department of Astronomy, University of Wisconsin, 475 North Charter Street, Madison, WI 53706, USA
* Hubble Fellow

**For decades we have known that the Sun lies within the Local Bubble, a cavity of low-density, high-temperature plasma surrounded by a shell of cold, neutral gas and dust.[1–3] However, the precise shape and extent of this shell[4,5], the impetus and timescale for its formation[6,7], and its relationship to nearby star formation[8] have remained uncertain, largely due to low-resolution models of the local interstellar medium. Leveraging new spatial[9–11] and dynamical constraints from the *Gaia* space mission[12], here we report an analysis of the 3D positions, shapes, and motions of dense gas and young stars within 200 pc of the Sun. We find that nearly all the star-forming complexes in the solar vicinity lie on the surface of the Local Bubble and that their young stars show outward expansion mainly perpendicular to the bubble's surface. Tracebacks of these young stars' motions support a scenario where the origin of the Local Bubble was a burst of stellar birth and then death (supernovae) taking place near the bubble's center beginning ~14 Myr ago. The expansion of the Local Bubble created by the supernovae swept up the ambient interstellar medium into an extended shell that has now fragmented and collapsed into the most prominent nearby molecular clouds, in turn providing robust observational support for the theory of supernova-driven star formation.**

In Figure 1a (interactive), we present a 3D map of the solar neighborhood, including a new *Gaia*-era 3D model of the Local Bubble's inner surface of neutral gas and dust[10,13] and the 3D shapes and positions of local molecular clouds constrained at ≈ 1 pc resolution[9,11]. The Local Bubble's shell is shown as a closed surface, but evidence suggests it could be a "Galactic chimney", having blown out of the disk a few hundred parsecs above and below the Galactic plane.[14] The distribution of dense gas in star-forming molecular clouds is shown with a set of topological "spines" derived by "skeletonizing" the clouds in 3D volume density space.[11]

Remarkably, we find that every well-known molecular cloud within ~200 pc of the Sun lies on the surface of the Local Bubble. These "surface" clouds include not just every star-forming region in the Scorpius-Centaurus (Sco-Cen) association (Ophiuchus, Lupus, Pipe, Chamaeleon, and Musca), but also the Corona Australis region and the Taurus Molecular Cloud, the latter of which lies 300 pc away from Sco-Cen on the opposite side of the bubble. The one exception is the Perseus Molecular Cloud, at a distance of 300 pc, which has likely been displaced by the recently discovered Per-Tau Superbubble[15], containing Taurus on its near side and Perseus on its far side (see green sphere in Figure 1b, interactive). The Taurus molecular cloud complex lies at the intersection of the Per-Tau Bubble and Local Bubble, displaying a sheet-like morphology consistent with being shaped by a bubble-bubble collision. Every Local Bubble surface cloud shows evidence of a similar sheet-like (e.g. Taurus) or filamentary (e.g. Corona Australis) morphology, uniformly elongated along the bubble's surface.

In Figure 1b, in addition to the Local Bubble surface, we show models for two kiloparsec-scale Galactic features discovered in the *Gaia* era: the Radcliffe Wave[16] and the Split[10]. The Radcliffe Wave is a 2.7-kpc-long filament of gas corresponding to the densest part of the Local Arm of the Milky Way. It has the shape of a damped sinusoid, extending above and below the plane of the Galaxy.[16] The Split, an at least 1-kpc-long gaseous feature situated in the disk, is argued to be a spur-like feature, bridging the Local and Sagittarius-Carina arms.[10] Also shown (interactive version only) is a model of the Gould's Belt, a disk of young stars, gas, and dust, tilted by about 20° with respect to the Galactic plane, which has long shaped our understanding of the architecture of the local interstellar medium. Prior work has suggested that the Gould's "Belt" is a superposition of unassociated structures seen in projection, with all well-known regions of the "Belt" being part of either the Radcliffe Wave or the Split large-scale gaseous structures.[16] As seen in Figure 1b (interactive), this argument about the nature of the Gould's Belt is confirmed here. The right side of the assumed Gould's Belt (the Sco-Cen association) consists of the entire rightward wall of the Local Bubble, while its left side consists of clouds in the Radcliffe Wave, well beyond the leftward wall of the Local Bubble. The Local Bubble lies at the closest distance between the Radcliffe Wave and the Split, with most of the dense gas at its surface co-spatial with these two kpc-scale features.

We use measurements of the 3D positions and motions of stellar clusters to reconstruct the star formation history near the Local Bubble. We rely on curated samples of young stars from the literature, as summarized in Extended Data Table 1. Our sample includes: clusters associated with star-forming regions on the surface of the bubble (Taurus, Ophiuchus, Lupus, Chamaeleon, and Corona Australis), older members of the Sco-Cen association (Upper Centaurus Lupus, Lower Centaurus Crux, and Upper Scorpius) up to a maximum age of 20 Myr; and clusters in known star-forming regions along the Radcliffe Wave and the Split but beyond the boundaries of the Local Bubble itself (Perseus, Serpens, Orion).

As described in the Methods section, we derive the "tracebacks" of stellar clusters associated with the Local Bubble and related structures. The current 3D space motions of the young stellar clusters are shown as cones in the interactive version of Figure 1, with the apex of the cone pointing in the direction of motion. Prior research has shown that the 3D space motions of the youngest clusters (≲ 3 Myr) can be considered probes of the 3D space motions of the parental gas clouds in which they were born.[17] Using the young stars' motions to trace cloud motion, we see that not only do all star-forming clouds presently observed within 200 pc *lie on the surface* of the Local Bubble, but they also show strong evidence of *outward expansion*, primarily perpendicular to the Bubble's surface. Tracebacks of the clusters' motions over the past 20 Myr point to the likely origin of the Local Bubble—presumably the region where the supernovae driving the bubble went off. The clear implication of the observed geometry and motions is that all the well-known star-forming regions within 200 pc of the Sun formed as gas has been swept up by the Local Bubble's expansion.

The interactive version of Figure 1 also includes a model for Gould's Belt[18], which illustrates that much of the motion previously attributed to the expansion of the assumed Gould's Belt[19] is instead likely due to the expansion of the Local Bubble. Recent work using complementary catalogs of young stars bolster this interpretation, finding evidence that the Sco-Cen stellar association — a key anchor of the Gould Belt — has a arc-like morphology consistent with recent sequential star formation, which we now know to be triggered by the Local Bubble.[20]

A full animation of the stellar tracebacks is provided in Figure 2 ([interactive](interactive)). In the static version, we show select snapshots at -16 Myr, -15 Myr, -14 Myr, -10 Myr, -6 Myr, -2 Myr, and the present. We observe multiple epochs of star formation, with each generation of stars consistent with being formed at the edge of the Local Bubble's expanding shell. We find that 15-16 Myr ago, the Upper Centaurus Lupus (UCL) and Lower Centaurus Crux (LCC) clusters in the Sco-Cen stellar association were born about 15 pc apart, and that the Bubble itself was likely created by supernovae whose surviving members belong to these clusters.

Based on the amount of momentum injection required by supernovae to sweep up the total mass of the shell ($1.4^{+0.65}_{-0.62} \times 10^6$ M☉) given its present-day expansion velocity ($6.7^{+0.5}_{-0.4}$ km/s), we estimate that $15^{+11}_{-7}$ supernovae were required to form the Local Bubble (see Methods section). Through an analysis of their existing stellar membership and an adopted Initial Mass Function (IMF), previous studies agree that UCL and LCC have produced 14-20 supernovae over their lifetimes.[6,7,21] However, previous work[6,7] also claims that UCL and LCC formed *outside* the present-day boundary of the Local Bubble, only entering its interior in the past few megayears, inconsistent with an argument that they are the progenitor population. By adopting new *Gaia* EDR3 estimates of the clusters' 3D velocities, better orbit integration, and a more accurate value for the Sun's peculiar motion, we find that UCL and LCC indeed coincide with the center of the bubble at its birth, lying just interior to its inner surface in the present day, thereby resolving this discrepancy. We explain the inconsistency between the stellar tracebacks for UCL and LCC proposed in this work and those from prior work in more detail in the Methods section.[6,7]

Under the assumption that each star-forming molecular cloud formed due to the shell's expansion – powered by UCL and LCC near its center – we fit for the temporal and radial evolution of the Local Bubble by building on recent analytic frameworks.[22] As described in the Methods section, our idealized, spherical shell expansion model fits for the age of the Local Bubble, the duration between supernova explosions powering its expansion, and the ambient density of the interstellar medium prior to the first explosion. We find that an age of $14.4^{+0.7}_{-0.8}$ Myr, a time between supernova explosions of $1.1^{+0.6}_{-0.4}$ Myr, and an ambient density of $2.7^{+1.6}_{-1.0}$ cm$^{-3}$ provides the best-fit to the dynamical tracebacks. This best-fit model for the Local Bubble's expansion is also shown in the static and [interactive](interactive) versions of Figure 2.

Following the presumed birth of the Local Bubble 14 Myr ago, we observe four subsequent epochs of star formation at the surface of its expanding shell, taking place ≈ 10 Myr ago, ≈ 6 Myr ago, ≈ 2 Myr ago, and the present day. Around 10 Myr ago, we observe the formation of the Upper Scorpius association, as well as an older, recently discovered companion stellar population in Ophiuchus. Next, 6 Myr ago, both Corona Australis and the older stellar population of Taurus were born. Around 2 Myr ago, we detect the birth of stars in Lupus and Chamaeleon, as well as the younger stellar populations of Taurus and Ophiuchus. Finally, in the present day, we observe the current distribution of dense star-forming molecular gas, enveloping the Local Bubble. In Figure 2, we also overlay the solar orbit, which indicates that the Sun only entered the bubble around 5 Myr ago, and that it was about 300 pc away when the first supernovae went off in UCL and LCC. If this expanding shell scenario is true, we would expect a total of $1.7^{+0.97}_{-0.63} \times 10^6$ M☉ of gas to be swept up by the Local Bubble over its lifetime, given our inferred ambient density of $2.7^{+1.6}_{-1.0}$ cm$^{-3}$ and the current radius of $165 \pm 6$ pc (see Methods section and Extended Data Figure 2). Based on the 3D dust currently enveloping the shell's surface, we obtain an actual swept-up mass of $1.4^{+0.65}_{-0.62} \times 10^6$ M☉, in agreement with the model estimate.

The circumstances that led to the birth of the progenitor populations UCL and LCC are more difficult to constrain. Given the close proximity of both the Radcliffe Wave and the Split to the Local Bubble, the origin of UCL and LCC could be related to one of these kpc-scale gaseous structures, or to a past interaction between the two. While current kinematic data are limited, the 3D tracebacks of young stars in two constituent clouds along the Radcliffe Wave (Orion) and the Split (Serpens), but beyond the Local Bubble's influence suggest that the Radcliffe Wave and the Split could have converged 20 Myr ago at the location where UCL and LCC were born. However, future follow-up work on the 3D motions of these two linear features would be needed to shed light on the true architecture of interstellar gas on kiloparsec scales at the time of UCL's and LCC's formation.

Regardless of UCL and LCC's potential origins, we find 6D (3D position and 3D velocity) observational support for the theory of supernova-driven star formation in the interstellar medium[23–26] − a long-invoked theoretical pathway for molecular cloud formation seen in numerical simulations.[27] The abundance of new stellar radial velocity data expected in *Gaia* DR3 should not only allow for more refined estimates of the Local Bubble's evolution, but also enable similar studies farther afield, providing further observational constraints on supernova-driven star formation across our Galactic neighborhood.

## References

1. Cox, D. P. & Reynolds, R. J. The local interstellar medium. *Annu. Rev. Astron. Astrophys.* **25**, 303–344 (1987).
2. Lucke, P. B. The distribution of color excesses and interstellar reddening material in the solar neighborhood. *Astron. Astrophys.* **64**, 367–377 (1978).
3. Sanders, W. T., Kraushaar, W. L., Nousek, J. A. & Fried, P. M. Soft diffuse X-rays in the southern galactic hemisphere. *Astrophys. J. Let.* **217**, L87–L91 (1977).

4. Lallement, R., Welsh, B. Y., Vergely, J. L., Crifo, F. & Sfeir, D. 3D mapping of the dense interstellar gas around the Local Bubble. *Astron. Astrophys.* **411**, 447–464 (2003).
5. Welsh, B. Y., Lallement, R., Vergely, J.-L. & Raimond, S. New 3D gas density maps of NaI and CaII interstellar absorption within 300 pc. *Astron. Astrophys.* **510**, A54 (2010).
6. Fuchs, B., Breitschwerdt, D., de Avillez, M. A., Dettbarn, C. & Flynn, C. The search for the origin of the Local Bubble redivivus. *Mon. Not. R. Astron. Soc.* **373**, 993–1003 (2006).
7. Breitschwerdt, D. *et al.* The locations of recent supernovae near the Sun from modelling $^{60}$Fe transport. *Nature.* **532**, 73–76 (2016).
8. Frisch, P. & Dwarkadas, V. V. Effect of Supernovae on the Local Interstellar Material. in *Handbook of Supernovae* (eds. Alsabti, A. W. & Murdin, P.) 2253–2285 (Springer International Publishing, 2017).
9. Leike, R. H., Glatzle, M. & Enßlin, T. A. Resolving nearby dust clouds. *Astron. Astrophys.* **639**, A138 (2020).
10. Lallement, R. *et al.* Gaia-2MASS 3D maps of Galactic interstellar dust within 3 kpc. *Astron. Astrophys.* **625**, A135 (2019).
11. Zucker, C. *et al.* On the Three-dimensional Structure of Local Molecular Clouds. *Astrophys. J.* **919**, 35 (2021).
12. Lindegren, L. *et al.* Gaia Early Data Release 3 - The astrometric solution. *Astron. Astrophys. Suppl. Ser.* **649**, A2 (2021).
13. Pelgrims, V., Ferrière, K., Boulanger, F., Lallement, R. & Montier, L. Modeling the magnetized Local Bubble from dust data. *Astron. Astrophys.* **636**, A17 (2020).
14. Welsh, B. Y., Sfeir, D. M., Sirk, M. M. & Lallement, R. EUV mapping of the local interstellar medium: the Local Chimney revealed? *Astron. Astrophys.* **352**, 308–316 (1999).
15. Bialy, S. *et al.* The Per-Tau Shell: A Giant Star-forming Spherical Shell Revealed by 3D Dust Observations. *Astrophys. J. Let.* **919**, L5 (2021).
16. Alves, J. *et al.* A Galactic-scale gas wave in the solar neighbourhood. *Nature.* **578**, 237–239 (2020).
17. Großschedl, J. E., Alves, J., Meingast, S. & Herbst-Kiss, G. 3D dynamics of the Orion cloud complex - Discovery of coherent radial gas motions at the 100-pc scale. *Astron. Astrophys. Suppl. Ser.* **647**, A91 (2021).
18. Perrot, C. A. & Grenier, I. A. 3D dynamical evolution of the interstellar gas in the Gould Belt. *Astron. Astrophys. Suppl. Ser.* **404**, 519–531 (2003).
19. Dzib, S. A., Loinard, L., Ortiz-León, G. N., Rodríguez, L. F. & Galli, P. A. B. Distances and Kinematics of Gould Belt Star-forming Regions with Gaia DR2 Results. *Astrophys. J.* **867**, 151 (2018).
20. Kerr, R. M. P., Rizzuto, A. C., Kraus, A. L. & Offner, S. S. R. Stars with Photometrically Young Gaia Luminosities Around the Solar System (SPYGLASS). I. Mapping Young Stellar Structures and Their Star Formation Histories. *Astrophys. J.* **917**, 23 (2021).
21. Maíz-Apellániz, J. The Origin of the Local Bubble. *Astrophys. J. Let.* **560**, L83–L86 (2001).
22. El-Badry, K., Ostriker, E. C., Kim, C.-G., Quataert, E. & Weisz, D. R. Evolution of supernovae-driven superbubbles with conduction and cooling. *Mon. Not. R. Astron. Soc.* **490**, 1961–1990 (2019).
23. Inutsuka, S.-I., Inoue, T., Iwasaki, K. & Hosokawa, T. The formation and destruction of molecular clouds and galactic star formation. An origin for the cloud mass function and star formation efficiency. *Astron. Astrophys.* **580**, A49 (2015).

24. Dawson, J. R. The Supershell–Molecular Cloud Connection: Large-Scale Stellar Feedback and the Formation of the Molecular ISM. *Publ. Astron. Soc. of Aust.* **30**, e025 (2013).
25. Cox, D. P. & Smith, B. W. Large-Scale Effects of Supernova Remnants on the Galaxy: Generation and Maintenance of a Hot Network of Tunnels. *Astrophys. J. Let.* **189**, L105–L108 (1974).
26. McKee, C. F. & Ostriker, J. P. A theory of the interstellar medium: three components regulated by supernova explosions in an inhomogeneous substrate. *Astrophys. J.* **218**, 148–169 (1977).
27. Kim, C.-G., Ostriker, E. C. & Raileanu, R. Superbubbles in the Multiphase ISM and the Loading of Galactic Winds. *Astrophys. J.* **834**, 25 (2017).

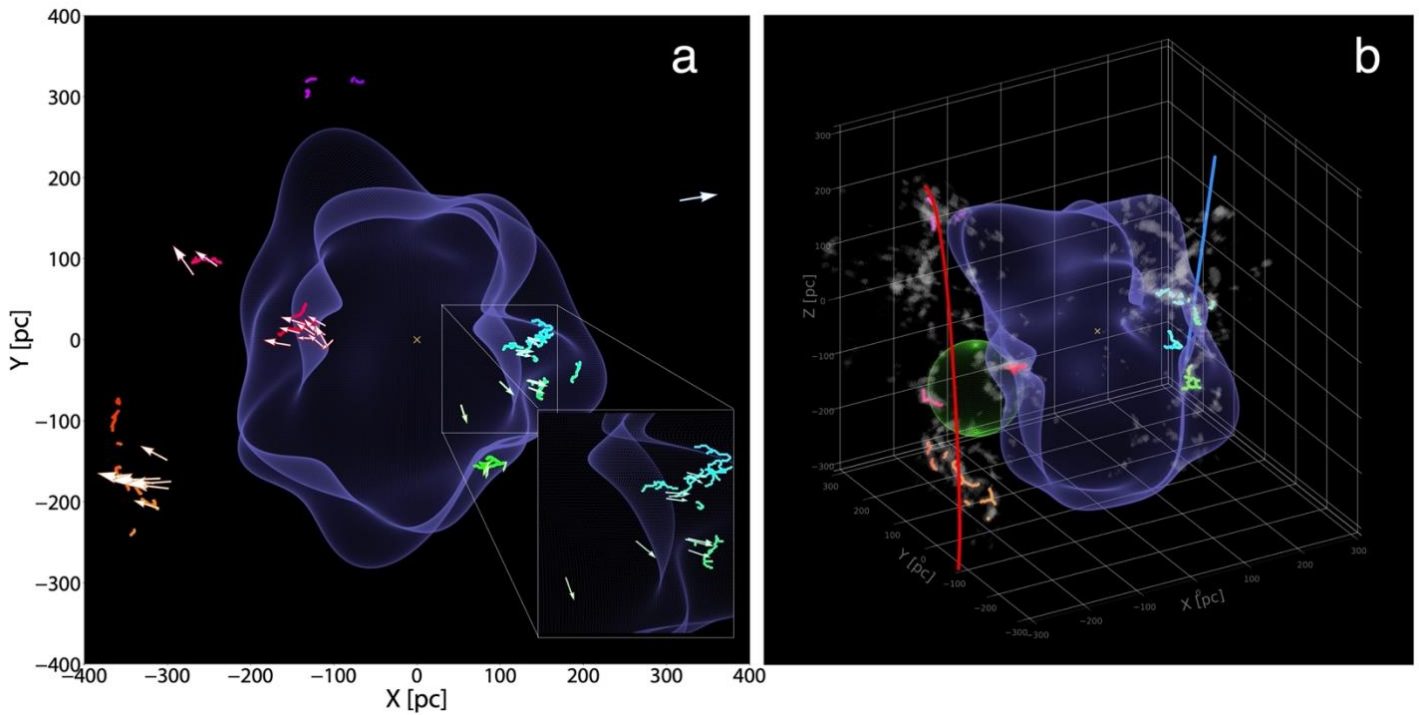

**Figure 1 (Interactive)**: **A 3D spatial view of the solar neighborhood.** For the best experience, please view the online 3D interactive version available here. **Panel a**: A top-down projection of star-forming regions on the surface of the Local Bubble, whose young stars show motion mainly perpendicular to its surface. The surface of the Local Bubble[13] is shown in purple. The short squiggly colored lines (a.k.a. "skeletons") demarcate the 3D spatial morphology of dense gas in prominent nearby molecular clouds[11]. The 3D arrows indicate the positions of young stellar clusters, with the apex of the arrow's cone pointing in the direction of stellar motion. Clusters are color-coded by longitude, as in Extended Data Table 1. The Sun is marked with a yellow cross. The zoom-in to the lower right shows a close-up of Ophiuchus, Pipe, Lupus, and Corona Australis on the Bubble's surface, along with arrows illustrating the outward motion of their young stellar clusters. **Panel b**: A 3D view of the relationship between the Local Bubble, prominent nearby star-forming regions, and Galactic structure. The Local Bubble and cloud skeletons are the same as in Panel a. We also overlay the morphology of the 3D dust (gray blobby shapes[9]) and the models for two Galactic scale features — the Radcliffe Wave (red)[16] and the Split (blue)[10]. The Per-Tau Superbubble[15] (green sphere) is also overlaid. The interactive version offers views from any direction (not just top-down), provides floating labels for star-forming regions, and includes additional layers (some not shown in this snapshot) which can be toggled on/off.

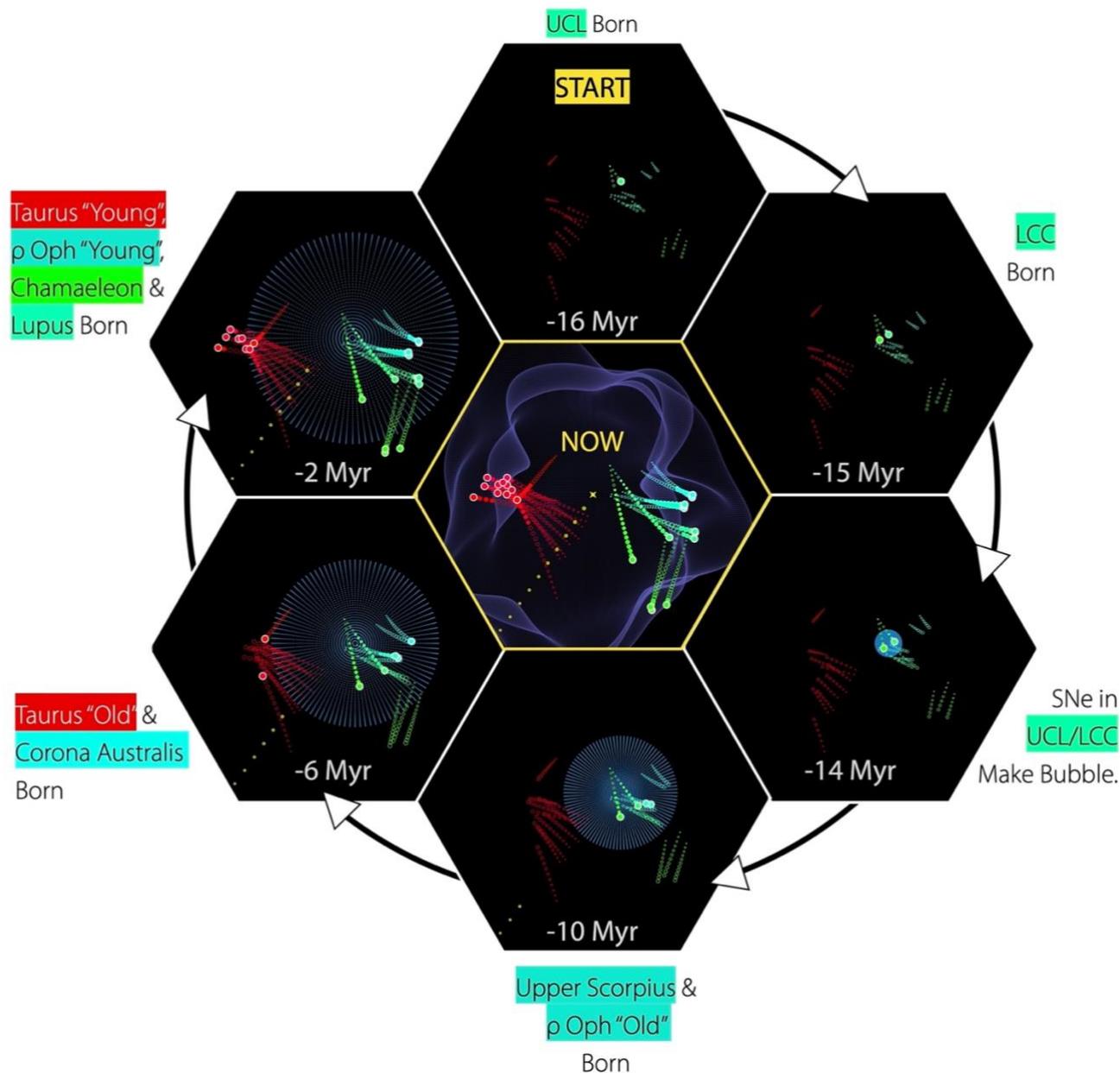

**Figure 2 (Interactive)**: **The evolution of the Local Bubble and sequential star formation at the surface of its expanding shell.** Selected time snapshots (seen from a top-down projection) are shown here. For a full time-sequence, viewable from any angle (not just top-down), see the online 3D interactive version available **here**. The central figure shows the present day. Stellar cluster tracebacks are shown with the colored paths. Prior to the cluster birth, the tracebacks are shown as unfilled circles meant to guide the eye, since our modeling is insensitive to the dynamics of the gas before its conversion into stars. After the cluster birth, the tracebacks are shown with filled circles and terminate in a large dot, which marks the cluster's current position. For time snapshots ≤ 14 Myr ago, we overlay a model for the evolution of the Local Bubble (purple sphere), as derived in the Methods section. The position of the local standard of rest (LSR) corresponds to the center of each panel. A more detailed overview of this evolutionary sequence, including the birth positions of all clusters, is provided in Extended Data Table 2. The solar orbit is shown in yellow and indicates that the Sun entered the Local Bubble approximately 5 Myr ago.

## Methods

**Deriving Stellar Cluster Properties**

In Extended Data Table 1, we summarize the properties of young stellar clusters utilized in this work. Our sample includes young clusters out to ≈ 300-400 pc, up to a maximum age of 20 Myr. Our sample is chosen to provide optimal coverage of all young (< 5 Myr) clusters associated with star-forming regions currently lying on the surface of the Local Bubble, including the Lupus[28], Ophiuchus[29], Chamaeleon[30], Corona Australis[31], and Taurus Molecular Clouds[32]. In addition to the youngest stellar populations on the Local Bubble's surface, several recent studies provide evidence that at least two of the surface clouds – Taurus[32] and Ophiuchus[29] – exhibit multiple generations of star formation, with an older stellar population (> 5 Myr) existing alongside the younger one. For these clouds, we include both young and old populations. Older stellar populations in the Scorpius Centaurus[33,34] association – Upper Centaurus Lupus, Lower Centaurus Crux, and Upper Scorpius are likewise included in Extended Data Table 1. While many of the known moving groups (e.g. Beta Pic, Octans, and Carina[33]) also lie inside the bubble, their ages (> 25 Myr) are larger than the bubble's estimated age and are thus excluded from our analysis[33]. Finally, we also include clusters beyond the Local Bubble but associated with star-forming regions along the Radcliffe Wave and the Split, including the Perseus[35], Orion[17], and Serpens[36] Molecular Clouds.

We rely on existing studies (see Extended Data Table 1) to determine stellar membership of each cluster, but we only include stars which are detected in the *Gaia* EDR3 catalog. We uniformly associate stellar members of each cluster with *Gaia* EDR3 using a crossmatch radius of 1" to obtain its sky coordinates, parallax, and proper motions. If radial velocities for the stars are provided along with the target selection in their original publication, we adopt those existing radial velocities in our analysis, some of which are obtained with high-quality ground based near-infrared spectroscopy. Otherwise, if no radial velocity data are provided, we utilize the *Gaia* radial velocities and restrict our analysis to only those stars which also have a *Gaia* radial velocity detection. We largely rely on the sample selection outlined in each cluster's original publication to filter outliers, many of which are defined using *Gaia* DR2 data. However, as an additional constraint, we require that all stars must have a parallax over error greater than two and small renormalized unit weight error (RUWE < 1.4).[37] We also require the radial velocity error to be < 5 km/s, a relatively generous cut chosen because our algorithm incorporates the errors on the individual stellar measurements when determining the mean cluster motion. After applying all these cuts, we perform a sigma-clipping procedure using the *astropy*[38] package, removing extreme outliers whose radial velocities are inconsistent with the rest of the cluster population at the $3\sigma$ level. Finally, we require that each cluster have at least three stellar members. The mean stellar membership is much higher than this, averaging 37 stars per cluster.

In order to transform the sky coordinates (right ascension $\alpha$ and declination $\delta$), parallax ($\pi$), proper motions ($\mu_{RA}, \mu_{DEC}$), and heliocentric radial velocities ($v_{helio}$) of members to an average 3D space position and 3D velocity of the cluster in a Heliocentric Galactic Cartesian reference frame (*x, y, z, U, V, W*), we utilize the extreme deconvolution algorithm.[39] The Extreme Deconvolution algorithm infers an n-dimensional distribution function using a Gaussian Mixture Model (GMM) given the presence of noisy, incomplete, and heterogeneous samples of the underlying population. We apply the algorithm to infer the average 6D phase information of each cluster, given the observed values and estimated error covariances of its stellar members. For each star we use the *astropy*[38] functionality within *galpy*[40] to compute the star's Heliocentric Galactic Cartesian coordinates ($x, y, z$) and associated space motions ($U, V, W$) given the observed *Gaia* quantities ($\alpha, \delta, \pi, \mu_{RA}, \mu_{DEC}, v_{helio}$). We assume $U$ points towards the Galactic center, $V$ points toward the direction of Galactic rotation, and $W$ points toward the North Galactic Pole. In order to accurately estimate errors on ($x, y, z, U, V, W$) for each star, we randomly sample one hundred times from a multi-dimensional Gaussian in ($\alpha, \delta, \pi, \mu_{RA}, \mu_{DEC}, v_{helio}$) space assuming Gaussian uncertainties on all parameters as reported in the *Gaia* EDR3 catalog. Transforming each sample to a Heliocentric Galactic Cartesian coordinate frame, we then calculate the covariance of the set of samples for each star. Feeding the set of ($x, y, z, U, V, W$) values for the individual stellar cluster members and their associated sample covariances into the Extreme Deconvolution algorithm, we obtain the mean and variance of a single 6D Gaussian defining the average 3D position and 3D space motion of each cluster, as shown in Extended Data Table 1. We adopt a peculiar solar motion of ($U_\odot, V_\odot, W_\odot$) = (10.0, 15.4, 7.8) km/s[41] and correct the ($U, V, W$) values of each cluster for this solar motion to obtain its current 3D space velocity with respect to the Local Standard of Rest (LSR) frame ($U_{LSR}, V_{LSR}, W_{LSR}$). We use the mean cluster motion to "traceback" its trajectory in the Galaxy. To "traceback" a cluster means to compute the 3D position and 3D motion that the cluster would have had at different times in the past, given estimates of its present 3D position and 3D motion. In practice, the full bound orbit of the cluster can be computed, and the small section of the orbit constituting its trajectory in the recent past is extracted.

We perform the dynamical traceback of each cluster using the *galpy* package[40], which supports orbit integrations in a Milky Way-like potential, consisting of a bulge, disk, and dark matter halo. We sample the orbit from -20 Myr to the present day. Alongside each cluster, we also trace the Sun's orbit backward in time over the past 20 Myr and correct each cluster's orbit for the Sun's peculiar motion to ensure all orbits remain in the Local Standard of Rest frame. We emphasize that *galpy* only accounts for the gravitational potential of the Galaxy and does not consider the gravitational effects of the parental gas clouds in which many of these clusters are embedded. However, given the large extent of the Local Bubble relative to any individual star-forming region, and the fact that the Galactic potential should dominate over the gravity of any local gas, *galpy* still serves as a useful probe of the dynamics of the Local Bubble. This argument is bolstered by recent results from numerical simulations, which indicate that stellar orbits can be recovered with high fidelity up to 20 Myr in the past, even without explicitly modeling non-axisymmetric components of the potential.[42] The dynamical tracebacks for all clusters in Extended Data Table 1 are publicly available at the Harvard Dataverse (https://doi.org/10.7910/DVN/E8PQOD).

**Modeling the Local Bubble's Expansion**

In this section, we derive an analytic model for the radius and expansion velocity of the Local Bubble as a function of time, using constraints provided by the dynamical traceback data summarized in Extended Data

Tables 1 and 2 (see the Data Availability section to access the full traceback data on each cluster). The results of this section underpin our model for the temporal evolution of the Local Bubble shown in the static and interactive versions of Figure 2.

To model the expansion of the Local Bubble, we utilize recent literature that leverages 1D spherically-symmetric hydrodynamic simulations using the Athena++ code to study the dynamical evolution of superbubbles driven by clustered supernovae in a uniform medium.[22] Specific treatment is given to the effects of cooling at the shell/bubble interface. Building on this recent literature, we adopt an analytic model[22] describing the radius, $R$, of the superbubble's expanding shell as a function of time $t$ since its birth, parameterized by the ambient density $n_0$ of the interstellar medium, the cooling efficiency at the shell's surface $\theta$, the time separation between supernovae explosions $\Delta t_{SNe}$ within the cluster powering its formation,, and the energy input per supernova explosion $E_{SN}$, as follows:

$$R(t) = 83 \text{ pc} \times (1-\theta)^{\frac{1}{5}} \left(\frac{E_{SN}}{10^{51} \text{ erg}}\right)^{\frac{1}{5}} \left(\frac{\Delta t_{SNe}}{0.1 \text{ Myr}}\right)^{\frac{-1}{5}} \left(\frac{n_0}{1 \text{ cm}^{-3}}\right)^{\frac{-1}{5}} \left(\frac{t}{1 \text{ Myr}}\right)^{\frac{3}{5}} \quad (1)$$

At any time $t$ the $(x, y, z)$ coordinates of the surface of this expanding spherical shell, centered on $(x_{cen}, y_{cen}, z_{cen})$, corresponding to the epicenter of the supernova explosions, can be parameterized as:

$$(x(t) - x_{cen})^2 + (y(t) - y_{cen})^2 + (z(t) - z_{cen})^2 = R(t)^2 \quad (2)$$

The 3D positions of each cluster at birth (derived from the dynamical tracebacks given the cluster's age) provide a constraint on the bubble's radius as a function of time. So, under the assumption that the formation of the young stars listed in Extended Data Table 2 was triggered by the shell's expansion, we can infer the parameters governing the Local Bubble's evolution using this analytic framework. We emphasize that this theoretical formalism is an approximation of the bubble's true morphology. We do not actually expect the bubble to expand spherically because the interstellar medium is highly turbulent with significant density fluctuations. Indeed the Local Bubble today is observed to have a complex, non-spherical morphology. Extended Data Table 2 lists the 3D $(x, y, z)$ birth positions for the subset of young clusters utilized to model the bubble expansion in the Local Standard of Rest Frame, given their stellar tracebacks and estimated ages.

Since several of the parameters governing the superbubble's evolution are degenerate, we make a number of simplifying assumptions. Prior work[6,7,21] estimates that 14-20 supernovae in Upper Centaurus Lupus (UCL) and Lower Centaurus Crux (LCC) stellar groups over the past ~13 Myr have together created the Local Bubble. The formalism of the analytic superbubble model assumes that all supernovae are driven from a single location. Rather than model each individual supernova explosion given the tracebacks of UCL and LCC, we assume that the epicenter of the explosion $(x_{cen}, y_{cen}, z_{cen})$ lay equidistant between UCL and LCC at the time of the first explosion, $t_{exp}$. This approximation is justified as UCL and LCC lay roughly co-spatial at early times, when the most powerful supernovae driving the superbubble's expansion would have gone off. We leave the time of the first supernova explosion, $t_{exp}$, as a free parameter in our model. The subsequent evolution of the

Local Bubble is governed by $t_{exp}$ and Equation (1). We assume a fixed energy input per supernova of $10^{51}$ erg. We also assume a fixed cooling efficiency $\theta$ of 0.7. However, we test a variety of cooling efficiencies, ranging from 0.4 - 0.9, and find that fixing the cooling efficiency to a value of 0.7 does not affect our estimate for the time of the first explosion, and only has a modest effect on the ambient density and duration between supernovae (with the variation falling within our reported uncertainties on these parameters, as we will later show in Extended Data Figure 1).

Having fixed the cooling efficiency and energy input per supernova, the free parameters of our model include the ambient density, $n_0$, the time between supernova explosions, $\Delta t_{SNe}$, and the time of the first supernova explosion $t_{exp}$. We infer the values of $n_0$, $\Delta t_{SNe}$, and $t_{exp}$ in a Bayesian framework. We assume that the density of the shell has a Gaussian profile with an uncertainty (or thickness) of $\Delta_R$, which corresponds to a log-likelihood of the following form:

$$\mathrm{Log}(L) = \frac{-1}{2} \sum_{i=1}^{n} \left[ \ln(2\pi \Delta_R^2) + \frac{[R(t_i, n_0, \Delta t_{SNe}) - r_i(t_i)]^2}{\Delta_R^2} \right] \quad (3)$$

Here, the $R(t_i, n_0, \Delta t_{SNe})$ term is the radius of the Local Bubble's expanding shell governed by Equations (1) and (2), evaluated at time $t_i$, corresponding to the difference between the time at which the $i$th cluster was born and the time of the first explosion ($t_i = t_{birth,i} - t_{exp}$). The Local Bubble's shell with radius $R(t_i)$ is centered on ($x_{cen}, y_{cen}, z_{cen}$), derived from the mean 3D position of UCL and LCC in the Local Standard of Rest at time $t_{exp}$, when the first supernova went off in UCL or LCC. The $r(t_i)$ term is the radius of a sphere with the same center as $R(t_i)$, such that the $i$th cluster born at traceback time $t_i$ lies on its surface, with coordinates of:

$$(x(t_i) - x_{cen})^2 + (y(t_i) - y_{cen})^2 + (z(t_i) - z_{cen})^2 = r(t_i)^2 \quad (4)$$

Finally, the $\Delta_R$ term in the log-likelihood given in Equation (3) should be interpreted as an error term, encompassing uncertainties in both the ages of the clusters and on their mean ($U_{LSR}, V_{LSR}, W_{LSR}$) motions. We infer $\Delta_R$ as an additional free parameter in our model. The total log-likelihood of all $n$ clusters is the sum of their individual log-likelihoods, evaluated at the respective time of their births, which will be optimized when the difference between $R(t_i)$ and $r(t_i)$ is minimized.

Using the log-likelihood in Equation (3), we sample for the values of $t_{exp}$, $n_0$, $\Delta t_{SNe}$, and $\Delta_R$ using the nested sampling code *dynesty*.[43] For $t_{exp}$, we adopt a truncated normal prior with a mean of -13 Myr and a standard deviation of 1 Myr over the range -16 Myr to -10 Myr (consistent with previous evolutionary synthesis models of UCL and LCC).[6,21] For $n_0$, we adopt a truncated log-normal prior with a mean of 2 cm$^{-3}$ and a standard deviation of a factor of two over the range 0.1 to 10 cm$^{-3}$, consistent with the density range explored in the Athena++ simulations underpinning the expansion model.[22] For $\Delta t_{SNe}$, we adopt a truncated log-normal prior with a mean of 0.8 Myr and a standard deviation of a factor of two, over the range 0.05 Myr to 3 Myr (consistent with previous estimates of approximately 16 supernovae occurring in UCL and LCC over the past 13 Myr)[7]. Finally, guided by the typical errors on the 3D motions (a few km/s) and the ages (a few Myr), we adopt a truncated normal prior on $\Delta_R$, with a mean of 15 pc and a standard deviation of 5 pc, over the range 0 to 30 pc. We run with the default parameters of *dynesty's* dynamic nested sampler.

The results of our sampling procedure are summarized in Extended Data Figure 1. We find a median value and 1σ errors (computed using the 16th, 50th, and 84th percentiles of the samples) of $\Delta t_{SNe} = 1.06^{+0.63}_{-0.39}$ Myr, $n_0 = 2.71^{+1.57}_{-1.02}$ cm$^{-3}$, $t_{exp} = -14.39^{+0.78}_{-0.74}$ Myr, and $\Delta_R = 23.31^{+2.54}_{-2.29}$ pc. The best-fit model corresponds to an epicenter of the bubble of $(x_{cen}, y_{cen}, z_{cen}) = (39, 7, -18)$ pc in the LSR frame 14.4 Myr ago. Adopting these median parameters, a model for the evolution of the Local Bubble is overlaid in Figure 2. We compute the radius of the Local Bubble's shell and its expansion velocity over its lifetime, as plotted in Extended Data Figure 2 (panels a. and b.). Based on our model, and leveraging the full set of posterior samples, we estimate a present-day expansion speed of $6.7^{+0.5}_{-0.4}$ km/s and a radius of $165 \pm 6$ pc. This present-day expansion speed is consistent with the current range of 3D velocity magnitudes of stars at the surface of its shell (≈ 5 - 9 km/s), assuming that the rest velocity of the shell lies within a few km/s of the Local Standard of Rest (as we expect it to, since the LSR is currently inside the Bubble).

However, as seen in Extended Data Figure 1, there is a strong covariance between $\Delta t_{SNe}$ and $n_0$ such that an increase in the density of the ambient medium can be compensated for by a decrease in the time between supernova explosions, and vice versa. Given the limitations of our modeling, we intend the superbubble evolution shown in Figure 2 to only serve as a possible, idealized, example of how the Local Bubble could have reached its present-day morphology.

**Potential Origin of the Local Bubble**

In this section, we seek to shed additional light on the origin of the Local Bubble, by using a momentum analysis to test whether UCL and LCC harbored enough supernovae to excavate a cavity the size of the Local Bubble. To do so, we compute the momentum of the Local Bubble's shell from its current expansion velocity (calculated in the previous section and shown in Extended Data Figure 2) and estimates of its mass (obtained from 3D dust mapping). We can then further build on the analytic superbubble model constrained above to obtain the expected average momentum injection per supernova, $\hat{p}$. The ratio of the total momentum of the shell to the average momentum injected per supernova provides a constraint on the number of supernovae required to power its expansion, which can then be compared with existing estimates for how many supernovae have gone off in UCL and LCC based on population synthesis modeling.[7,21]

To obtain the momentum of the shell, we calculate its swept-up mass $M_{shell}$ by integrating the 3D volume density derived from the 3D dust map[9] between a distance of [$R_{shell}$, $R_{shell} + R_{thickness}$] from the Sun and multiplying by the mean particle mass $\langle m \rangle = 1.4 m_H$, where $m_H$ is the mass of a proton, and the factor of 1.4 corrects for the helium abundance. $R_{shell}$ is the boundary (i.e. inner radius) of the Local Bubble (shown in Figure 1) and $R_{thickness}$ is the shell's thickness, such that $R_{shell} + R_{thickness}$ corresponds to the outer radius. The Local Bubble model we utilize estimates $R_{thickness}$ to be between 50 - 150 pc[13], which is quite large, but encompasses the full depth of structure currently lying on the bubble' surface.

Adopting $R_{thickness}$ = 100 pc with an estimated $1\sigma$ uncertainty of 50 pc, we obtain $M_{shell} = 1.4^{+0.65}_{-0.62} \times 10^6$ M⊙ for the swept-up mass. To estimate the current expansion velocity of the Bubble $v_{exp}$, we leverage the posterior samples from our model describing the Local Bubble's evolution (fit to the dynamical tracebacks; see Extended Data Figure 1) to evaluate the velocity of the shell $v_{shell} = dR/dt$ in the present day (see Extended Data Figure 2). Doing so, we obtain $v_{exp} = 6.7^{+0.5}_{-0.4}$ km/s. To propagate uncertainties throughout this section, we use the full set of posterior samples when leveraging parameters from the expansion model. To propagate uncertainties on the observed swept up mass of the shell, we draw the same number of samples from a Gaussian with a mean of $M_{shell}$ = 1.4 x 10⁶ M⊙ and a standard deviation of 6 x 10⁵ M⊙. Whenever uncertainties are reported, we use the 16th, 50th, and 84th percentiles of the samples to compute the median and $1\sigma$ error bounds. Given samples for the current expansion velocity and swept-up mass, the corresponding momentum of the Local Bubble's shell is:

$$p_{shell} = M_{shell} \times v_{exp} = 9.6^{+4.4}_{-4.1} \times 10^6 \text{ M⊙ km/s}$$

In the previous section, we utilized an analytic superbubble expansion model[22] to constrain the temporal evolution of the Local Bubble's size and velocity. The same simulations underpinning this formalism indicate that the average momentum injected per supernovae $\hat{p}$ depends on the same free parameters, namely the energy input per supernova explosion, $E_{SN}$, the cooling efficiency, $\theta$, the duration between supernova explosions, $\Delta t_{SNe}$, the density of the ambient medium, $n_0$, and time $t$:

$$\hat{p}(t) = 4 \times 10^5 \, M_\odot \text{ km s}^{-1} (1-\theta)^{\frac{4}{5}} \left(\frac{E_{SN}}{10^{51} \text{ erg}}\right)^{\frac{4}{5}} \left(\frac{\Delta t_{SNe}}{0.1 \text{ Myr}}\right)^{\frac{1}{5}} \left(\frac{n_0}{1 \text{ cm}^{-3}}\right)^{\frac{1}{5}} \left(\frac{t}{1 \text{ Myr}}\right)^{\frac{2}{5}} \quad (5)$$

Therefore, we again fix $\theta$ = 0.7, $E_{SN}$ = 10⁵¹ erg, and leverage the samples of $\Delta t_{SNe}$, $n_0$, and $t_{exp}$. We evaluate $\hat{p}$ over the lifetime of each realization of the bubble and draw a random sample of $\hat{p}$ from each distribution. We then take the 16th, 50th, and 84th percentiles of the random samples drawn from all realizations (see Extended Data Figure 1) to obtain a mean value and $1\sigma$ uncertainties on the average momentum injected per supernova $\hat{p}$:

$$\hat{p} = 6.5^{+1.6}_{-2.4} \times 10^5 \text{ M⊙ km/s}$$

Finally, dividing all samples of $p_{shell}$ by all samples of $\hat{p}$ we obtain:

$$N_{SNe} = p_{shell} / \hat{p} = 15^{+11}_{-7} \text{ supernovae}$$

The marginal distribution of $N_{SNe}$ is shown in Extended Data Figure 3. The average number of $15^{+11}_{-7}$ supernovae is in good agreement with previous results[6,7,21], which argue that UCL and LCC have produced 14-20 supernovae over their lifetimes based on an analysis of their current stellar membership and modeling of a Salpeter[44] IMF.

Consistent with this physical scenario, before supernovae start going off in UCL and LCL, our model predicts a typical ambient density of the interstellar medium of $n_0 = 2.71^{+1.57}_{-1.02}$ cm$^{-3}$. Assuming that the volume of sphere with a radius of $165 \pm 6$ pc (our current estimate for the radius of the Local Bubble from the analytic expansion model) is uniformly filled with a gas density of $n_0 = 2.71^{+1.57}_{-1.02}$ cm$^{-3}$, we would expect $1.7^{+0.97}_{-0.63} \times 10^6$ M⊙ of gas to be displaced and swept up into its surrounding shell over its lifetime. Using the 3D dust measurements, we measure an actual swept-up mass of $M_{\text{shell}} = 1.4^{+0.65}_{-0.62} \times 10^6$ M⊙ on the surface of the Local Bubble, aligned with this estimate of 1.7 million M⊙.

Given uncertainties in the exact volume density of gas and dust inside the bubble traced by 3D dust – with some recent work suggesting that there is only a modest drop in volume density coincident with the cavity of hot ionized gas[45] – we note that these momentum calculations (and more broadly the analytic superbubble expansion model) are largely insensitive to the exact difference in density interior and exterior to the bubble's boundary. The analytic expansion model we adopt[22] includes the ambient density of the interstellar medium at the time of the first supernova explosion as a free parameter; however, this ambient density parameter is constrained using the dynamical traceback data (not the 3D dust maps) and does not require any explicit assumptions about the exact density interior and exterior to the bubble as the shell expands. Similarly, while the swept-up mass calculation does depend on 3D dust mapping[9], the mass is only measured within the denser shell of neutral gas and dust (where the uncertainties on the underlying dust extinction are smaller) and has no dependence on the density in the hot inner cavity, which could be subject to larger uncertainties due to the very low levels of dust extinction.

**The Stellar Tracebacks of UCL/LCC and the Peculiar Motion of the Sun**

Tracebacks of the average stellar cluster motions are expressed relative to the "Local Standard of Rest," which itself depends on the measured motion of the Sun. Even in the age of *Gaia*, the Sun's exact motion in the Galaxy is uncertain, especially along the "Y" direction (along Galactic rotation). In this paper, we revise estimates of the tracebacks of the UCL and LCC clusters from prior literature[6,7], which argues that UCL and LCC were born *outside* of the present-day boundary of the Local Bubble even though their members were the likely *progenitors* of the Local Bubble. In this section, we revisit these extant stellar traceback calculations from prior work[6,7] finding that a revised calculation places UCL and LCC near the *center* of the Local Bubble when they formed.

While previous studies[21] have proposed UCL and LCC as the likely progenitor population, the prior literature in question[6] is the first study to perform an unbiased stellar search of all nearby B stars to track down the remains of OB associations hosting supernovae capable of powering the Local Bubble. This prior literature[6] confirms that UCL and LCC are the only populations capable of having powered the Local Bubble, but, after tracing back the stellar members, find that both UCL and LCC formed > *100 pc outside* the present-day boundary of the Local Bubble. Extended Data Figure 4a shows the extant stellar traceback results, where UCL and LCC only lay interior to the present-day boundary of Local Bubble's during the past few megayears.[6] The prior literature[6] noted this potential inconsistency (how does a supernova cause a bubble it does not lie within?), but argue that the location of the UCL and LCC clusters with respect to the center of the bubble is not crucial.

The prior literature[6] which places UCL and LCC outside the bubble use Hipparcos parallaxes and proper motions with radial velocities collected from the literature to derive the (U, V, W) 3D space motion of each member. Using the (U, V, W) value of each individual star, prior literature[6] calculates the mass–weighted mean (U, V, W) velocity of all members of UCL and LCC, and then assign this group 3D U, V, W space velocity to each individual stellar member for the purposes of performing the tracebacks. The prior literature[6] also states that they correct the mean 3D group velocity for the Sun's peculiar motion, adopting a value of ($U_\odot$, $V_\odot$, $W_\odot$) = (10.0, 5.2, 7.2) km/s[46], so that the traceback of each star is reported in the Local Standard of Rest frame.

As a first step, we attempt to reproduce the original results from prior literature[6] (see Extended Data Figure 4a) using their own data. Specifically, we calculate the mass-weighted mean velocity of the UCL and LCC stars from their Table A1[6] and Equation 2[6], finding (U, V, W = -7.1 -20.6, -5.8) km/s without correcting for the solar motion. With a value for the Sun's peculiar motion of (10.0, 5.2, 7.2) km/s[46] used in the prior literature, these values translate to ($U_{LSR}$, $V_{LSR}$, $W_{LSR}$ = 2.9, -15.4, 1.4) km/s in the LSR frame. Extended Data Figure 4b shows dynamical tracebacks we calculate in *galpy* (see Methods) from the Table A1 data in prior literature and with the value for the Sun's peculiar motion of (10.0, 5.2, 7.2) km/s[46]. Using the default orbit integrator in *galpy* with its standard Milky Way potential[40], we are *unable to reproduce the results from prior literature*, particularly the strong curvature in the tracebacks along the +X direction. The prior literature uses an epicyclic approximation for the stars' motions (see their Section 2)[6], which may be responsible for part of the discrepancy. This systematic smearing out of the tracebacks toward +X in their Figure 2[6] manifests in the entire sample of B stars and is not isolated to UCL and LCC.

In examining the stellar motions of UCL and LCC, we find that the choice of the solar peculiar motion – necessary to convert to the LSR frame – has a non-negligible effect on where the birthplaces of UCL and LCC lie with respect to the center of the Local Bubble. There have been dozens of attempts to constrain the Sun's peculiar motion, but $V_\odot$ (motion in the direction of Galactic rotation, along Y) remains highly uncertain: current estimates range between $V_\odot$ = 4 and 16 km/s.[47]

The value of $V_\odot$ = 5.2 km/s[46] adopted in prior literature[6,7] is one of the lowest values of $V_\odot$. As Extended Data Figure 4c shows, if we use the same Hipparcos data utilized in the prior literature but replace their value for $V_\odot$ (5.2 km/s[46]) with a value of 15.4 km/s[41] we find that UCL and LCC are born *near the bubble's center*. Extended Data Figure 4d shows that updating the Hipparcos data with *Gaia* data and adopting the same peculiar motion used in this work[41] makes almost no difference to the tracebacks, suggesting that uncertainty in Sun's peculiar motion is the dominant source of uncertainty in determining the birth location of UCL and LCC. We adopt a value of 15.4 km/s[41] when calculating all the tracebacks in this paper, toward the upper end of estimates for $V_\odot$. Given the large uncertainty on $V_\odot$, we have tested the robustness of our results to the choice of solar peculiar motion and find that any motion for $V_\odot \gtrsim$ 10 km/s is entirely consistent with the physical scenario we propose here. This $V_\odot$ = 10 to 16 km/s range encompasses the vast majority of estimates for $V_\odot$ used in the field today.[47–50]

**Data Availability**

The datasets generated and/or analyzed during the current study are publicly available on the Harvard Dataverse (https://dataverse.harvard.edu/dataverse/local_bubble_star_formation/), including Extended Data Table 1 (doi:10.7910/DVN/ZU97QD), Extended Data Table 2 (doi:10.7910/DVN/1VT8BC), per-star data for individual stellar cluster members (doi:10.7910/DVN/1UPMDX), and the cluster tracebacks (doi:10.7910/DVN/E8PQOD).

**Code Availability**

The results generated in this work are based on publicly available software packages and do not involve the extensive use of custom code. Given each star's reported *Gaia* data, we use the *astropy*[38] package to obtain the Heliocentric Galactic Cartesian positions and velocities. The extreme deconvolution algorithm in the *astroML*[51] package is used to estimate the mean 3D positions and velocities of the stellar clusters. The *Orbit* functionality in the *galpy*[40] package is used to perform the dynamical tracebacks. The *dynesty*[43] package is used to fit the analytic superbubble expansion model and determine the best-fit parameters governing the Local Bubble's evolution.

**Methods References**

28. Galli, P. A. B. *et al.* Lupus DANCe. Census of stars and 6D structure with Gaia-DR2 data. *Astron. Astrophys.* **643**, A148 (2020).
29. Grasser, N. *et al.* The ϱ Oph region revisited with Gaia EDR3. *Astron. Astrophys.* **652**, A2 (2021)
30. Galli, P. A. B. *et al.* Chamaeleon DANCe. Revisiting the stellar populations of Chamaeleon I and Chamaeleon II with Gaia-DR2 data. *Astron. Astrophys.* **646**, A46 (2021).
31. Galli, P. A. B. *et al.* Corona-Australis DANCe. I. Revisiting the census of stars with Gaia-DR2 data. *Astron. Astrophys.* **634**, A98 (2020).
32. Krolikowski, D. M., Kraus, A. L. & Rizzuto, A. C. Gaia EDR3 Reveals the Substructure and Complicated Star Formation History of the Greater Taurus-Auriga Star Forming Complex. arXiv:2105.13370 (2021).
33. Gagné, J. & Faherty, J. K. BANYAN. XIII. A First Look at Nearby Young Associations with Gaia Data Release 2. *Astrophys. J.* **862**, 138 (2018).
34. Gagné, J. *et al.* BANYAN. XI. The BANYAN Σ Multivariate Bayesian Algorithm to Identify Members of Young Associations with 150 pc. *Astrophys. J.* **856**, 23 (2018).
35. Ortiz-León, G. N. *et al.* The Gould's Belt Distances Survey (GOBELINS). V. Distances and Kinematics of the Perseus Molecular Cloud. *Astrophys. J.* 865, 73 (2018).
36. Herczeg, G. J. *et al.* An Initial Overview of the Extent and Structure of Recent Star Formation within the Serpens Molecular Cloud Using Gaia Data Release 2. *Astrophys. J.* **878**, 111 (2019).
37. Fabricius, C. *et al.* Gaia Early Data Release 3 - Catalogue validation. *Astron. Astrophys. Suppl. Ser.* **649**, A5 (2021).
38. The Astropy Collaboration *et al.* The Astropy Project: Building an Open-science Project and Status of the v2.0 Core Package*. *Astron. J. Supp.* **156**, 123 (2018).
39. Bovy, J., Hogg, D. W. & Roweis, S. T. Extreme deconvolution: Inferring complete distribution functions from noisy, heterogeneous and incomplete observations. *Ann. Appl. Stat.* **5**, 1657–1677 (2011).

40. Bovy, J. galpy: A python Library for Galactic Dynamics. *Astrophys. J. Supp.* **216**, 29 (2015).
41. Kerr, F. J. & Lynden-Bell, D. Review of galactic constants. *Mon. Not. R. Astron. Soc.* **221**, 1023–1038 (1986).
42. Kamdar, H., Conroy, C. & Ting, Y.-S. Stellar Streams in the Galactic Disk: Predicted Lifetimes and Their Utility in Measuring the Galactic Potential. *arXiv:2106.02050* (2021).
43. Speagle, J. S. dynesty: a dynamic nested sampling package for estimating Bayesian posteriors and evidences. *Mon. Not. R. Astron. Soc.* **493**, 3132–3158 (2020).
44. Salpeter, E. E. The luminosity function and stellar evolution. *Astrophys. J.* (1955).
45. Gontcharov, G. & Mosenkov, A. Interstellar polarization and extinction in the Local Bubble and the Gould Belt. *Mon. Not. R. Astron. Soc.* 483, 299-314 (2019).
46. Dehnen, W. & Binney, J. J. Local stellar kinematics from Hipparcos data. *Mon. Not. R. Astron. Soc.* **298**, 387–394 (1998).
47. Francis, C. & Anderson, E. Calculation of the local standard of rest from 20574 local stars in the New Hipparcos Reduction with known radial velocities. *New Astron.* **14**, 615–629 (2009).
48. Wang, F. *et al.* Local stellar kinematics and Oort constants from the LAMOST A-type stars. *Mon. Not. R. Astron. Soc.* **504**, 199–207 (2021).
49. Reid, M. J. *et al.* Trigonometric Parallaxes of High-mass Star-forming Regions: Our View of the Milky Way. *Astrophys. J.* **885**, 131 (2019).
50. Schönrich, R., Binney, J. & Dehnen, W. Local kinematics and the local standard of rest. *Mon. Not. R. Astron. Soc.* **403**, 1829–1833 (2010).
51. VanderPlas, J., Connolly, A. J., Ivezić, Ž. & Gray, A. Introduction to astroML: Machine learning for astrophysics. in *2012 Conference on Intelligent Data Understanding* 47–54 (2012).

## Acknowledgements

The visualization, exploration, and interpretation of data presented in this work were made possible using the glue visualization software, supported under NSF grant numbers OAC-1739657 and CDS&E:AAG-1908419. The interactive figures were made possible by the plot.ly python library. DPF acknowledges support by NSF grant AST-1614941 "Exploring the Galaxy: 3-Dimensional Structure and Stellar Streams." DPF, and AAG acknowledge support by NASA ADAP grant 80NSSC21K0634 "Knitting Together the Milky Way: An Integrated Model of the Galaxy's Stars, Gas, and Dust." AB acknowledges support by the Excellence Cluster ORIGINS which is funded by the German Research Foundation (DFG) under Germany's Excellence Strategy - EXC-2094-390783311. JA acknowledges support from the Data Science Research Center and the TURIS Research Platform of the University of Vienna. JG acknowledges funding by the Austrian Research Promotion Agency (FFG) under project number 873708. CZ acknowledges that support for this work was provided by NASA through the NASA Hubble Fellowship grant #HST-HF2-51498.001 awarded by the Space Telescope Science Institute, which is operated by the Association of Universities for Research in Astronomy, Inc., for NASA, under contract NAS5-26555. CZ, AAG, JA, and SB acknowledge Interstellar Institute's program "The Grand Cascade" and the Paris-Saclay University's Institut Pascal for hosting discussions that nourished the development of the ideas behind this work.

## Author Contributions

CZ led the work and wrote the majority of the text. All authors contributed to the text. CZ, AAG, and JA led interpretation of the observational results, aided by SB, MF, and AB who helped interpret their significance in light of theoretical models for supernova-driven star formation. CZ and AAG led the visualization efforts. JSS and DPF helped shape the statistical modeling of the Local Bubble's expansion. CZ, AAG, and JSS contributed to the software used in this work. JG and CS provided data for and the subsequent interpretation of the 3D kinematics of the Orion region. DK helped to develop the code used to model the 3D positions and motions of stellar clusters described in the Methods section.

## Competing Interests

The authors declare that they have no competing financial interests.

## Author Information

Correspondence and requests for materials should be addressed to CZ (email: czucker@stsci.edu or catherine.zucker@cfa.harvard.edu). CZ is a NASA Hubble Fellow at the Space Telescope Science Institute.

| Cluster Region | Cluster Subgroup | Sample | N_stars | l ° | b ° | d pc | x pc | σ_x pc | y pc | σ_y pc | z pc | σ_z pc | U_LSR km/s | σ_ULSR km/s | V_LSR km/s | σ_VLSR km/s | W_LSR km/s | σ_WLSR km/s | Age Myr |
|---|---|---|---|---|---|---|---|---|---|---|---|---|---|---|---|---|---|---|---|
| (1) | (2) | (3) | (4) | (5) | (6) | (7) | (8) | (9) | (10) | (11) | (12) | (13) | (14) | (15) | (16) | (17) | (18) | (19) | (20) |
| Perseus | NGC1333 | Ortiz-Leon et al. 2018 | 28 | 158.3 | -20.5 | 293 | -255 | 9 | 101 | 3 | -102 | 3 | -7.2 | 0.9 | 5.0 | 1.2 | -2.1 | 0.9 | 1 |
| Perseus | IC348 | Ortiz-Leon et al. 2018 | 126 | 160.5 | -17.8 | 315 | -282 | 8 | 100 | 3 | -96 | 3 | -6.9 | 1.0 | 9.6 | 0.9 | 0.1 | 0.9 | 3 |
| Taurus | C2 - L1495 | Krolikowski et al. 2021 | 24 | 168.9 | -15.8 | 130 | -122 | 2 | 24 | 1 | -35 | 2 | -6.0 | 0.5 | 3.2 | 0.7 | -3.1 | 0.5 | 1 |
| Taurus | C8 - B213 | Krolikowski et al. 2021 | 12 | 170.9 | -15.8 | 159 | -151 | 4 | 24 | 2 | -43 | 2 | -7.9 | 0.3 | 2.4 | 0.5 | 0.7 | 0.2 | 3 |
| Taurus | D4 - North | Krolikowski et al. 2021 | 21 | 172.3 | -16.3 | 135 | -128 | 25 | 17 | 7 | -38 | 9 | -4.9 | 2.9 | 5.2 | 5.9 | -1.4 | 3.1 | 2 |
| Taurus | C6 - L1524 | Krolikowski et al. 2021 | 32 | 173.7 | -15.5 | 128 | -123 | 2 | 13 | 4 | -34 | 1 | -5.8 | 0.9 | 3.6 | 1.6 | -1.9 | 0.9 | 2 |
| Taurus | C7 - L1527 | Krolikowski et al. 2021 | 11 | 174.0 | -13.6 | 140 | -136 | 3 | 14 | 1 | -33 | 1 | -5.9 | 1.3 | 4.0 | 0.7 | -2.2 | 0.7 | 3 |
| Taurus | C5 - L1546 | Krolikowski et al. 2021 | 13 | 175.6 | -16.5 | 162 | -155 | 3 | 12 | 1 | -45 | 1 | -6.7 | 0.5 | 1.7 | 0.4 | 0.6 | 0.7 | 2 |
| Taurus | D3 - South | Krolikowski et al. 2021 | 13 | 176.7 | -18.0 | 122 | -116 | 3 | 6 | 10 | -37 | 7 | -4.1 | 1.8 | 7.6 | 2.6 | -2.0 | 0.8 | 6 |
| Taurus | C1 - L1551 | Krolikowski et al. 2021 | 14 | 179.1 | -20.0 | 145 | -136 | 2 | 2 | 1 | -49 | 1 | -5.1 | 1.7 | 0.3 | 0.7 | 0.5 | 1.1 | 2 |
| Taurus | D2 - L1558 | Krolikowski et al. 2021 | 13 | 179.9 | -20.0 | 131 | -123 | 29 | 0 | 8 | -45 | 11 | -5.0 | 2.5 | 4.3 | 5.2 | -1.4 | 1.4 | 3 |
| Taurus | D1 - L1544 | Krolikowski et al. 2021 | 12 | 181.3 | -7.3 | 172 | -170 | 8 | -3 | 5 | -21 | 6 | -8.9 | 0.6 | 1.5 | 1.7 | -1.0 | 1.2 | 3 |
| Taurus | C9 - 118TauEast | Krolikowski et al. 2021 | 4 | 184.0 | -4.9 | 109 | -109 | 2 | -7 | 1 | -9 | 2 | -3.4 | 0.4 | -3.4 | 0.2 | -1.2 | 0.1 | 6 |
| Orion | L1616 | Großschedl et al. 2021 | 5 | 203.6 | -24.6 | 382 | -318 | 15 | -139 | 6 | -159 | 6 | -8.8 | 2.7 | 4.7 | 1.1 | -1.6 | 1.0 | 2 |
| Orion | NGC2068/2071 | Großschedl et al. 2021 | 44 | 205.2 | -14.3 | 413 | -362 | 10 | -170 | 5 | -102 | 3 | -13.6 | 1.3 | 3.2 | 1.0 | -0.5 | 1.3 | 2 |
| Orion | NGC 2023/2024 | Großschedl et al. 2021 | 37 | 206.6 | -16.3 | 401 | -344 | 10 | -172 | 5 | -112 | 3 | -13.5 | 1.8 | 2.6 | 1.6 | 0.5 | 1.2 | 2 |
| Orion | Sigma Ori | Großschedl et al. 2021 | 38 | 206.8 | -17.3 | 401 | -342 | 7 | -173 | 4 | -119 | 3 | -16.3 | 1.0 | -0.5 | 0.8 | 0.4 | 1.1 | 3 |
| Orion | NGC1977 | Großschedl et al. 2021 | 42 | 208.5 | -19.1 | 393 | -326 | 6 | -177 | 3 | -128 | 2 | -15.1 | 1.3 | -0.9 | 1.0 | -1.1 | 1.1 | 3 |
| Orion | Orion A, Head | Großschedl et al. 2021 | 241 | 209.2 | -19.5 | 393 | -324 | 9 | -181 | 5 | -131 | 4 | -12.2 | 2.0 | 1.6 | 1.9 | 0.8 | 1.6 | 2 |
| Orion | Orion A, Tail | Großschedl et al. 2021 | 153 | 211.8 | -19.4 | 408 | -327 | 17 | -203 | 19 | -135 | 8 | -8.2 | 2.7 | 2.8 | 1.5 | 0.5 | 1.0 | 2 |
| Chamaeleon | 1-North | Galli et al. 2021 | 34 | 297.0 | -14.8 | 192 | 84 | 1 | -165 | 2 | -49 | 1 | -0.6 | 0.6 | -3.8 | 0.7 | -3.9 | 0.6 | 2 |
| Chamaeleon | 1-South | Galli et al. 2021 | 34 | 297.2 | -15.8 | 188 | 82 | 2 | -161 | 3 | -51 | 2 | -1.6 | 0.6 | -4.2 | 1.1 | -3.4 | 0.6 | 2 |
| Sco-Cen | Lower Centaurus Crux, LCC | Gagne et al. 2018a,b | 42 | 301.3 | 6.0 | 110 | 57 | 14 | -93 | 13 | 11 | 15 | 1.9 | 1.5 | -6.1 | 2.0 | 1.9 | 1.6 | 15 |
| Chamaeleon | 2 | Galli et al. 2021 | 7 | 303.7 | -14.4 | 196 | 105 | 2 | -158 | 2 | -48 | 2 | -0.8 | 1.9 | -3.2 | 3.3 | -0.8 | 1.4 | 2 |
| Sco-Cen | Upper Centaurus Lupus, UCL | Gagne et al. 2018a,b | 40 | 330.7 | 13.3 | 129 | 109 | 25 | -61 | 17 | 29 | 16 | 4.8 | 3.5 | -4.6 | 1.9 | 2.7 | 1.4 | 16 |
| Lupus | 4 | Galli et al. 2020 | 5 | 336.4 | 8.4 | 160 | 145 | 1 | -63 | 1 | 23 | 1 | 5.2 | 1.5 | -2.0 | 0.9 | 0.2 | 0.3 | 3 |
| Lupus | 3 | Galli et al. 2020 | 27 | 339.5 | 9.4 | 158 | 146 | 2 | -54 | 1 | 25 | 1 | 6.7 | 2.7 | -2.0 | 1.3 | 0.2 | 0.9 | 3 |
| Lupus | Off-Cloud Population | Galli et al. 2020 | 3 | 340.1 | 10.9 | 158 | 146 | 4 | -52 | 1 | 29 | 5 | 5.8 | 3.6 | -1.4 | 1.2 | 0.2 | 0.5 | 3 |
| Sco-Cen | Upper Scorpius, USCO | Gagne et al. 2018a,b | 43 | 351.4 | 22.0 | 142 | 130 | 8 | -19 | 7 | 53 | 9 | 5.1 | 2.1 | -0.8 | 1.2 | 0.9 | 1.7 | 10 |
| Ophiuchus | Rho Oph, Population I | Grasser et al. 2021 | 96 | 353.1 | 17.2 | 138 | 131 | 3 | -15 | 1 | 40 | 2 | 4.7 | 1.7 | 0.3 | 0.9 | -1.4 | 1.0 | 1 |
| Ophiuchus | Rho Oph, Population II | Grasser et al. 2021 | 32 | 353.9 | 18.3 | 139 | 131 | 6 | -14 | 2 | 43 | 3 | 5.9 | 2.2 | -0.4 | 1.3 | 1.8 | 0.8 | 10 |
| Corona Australis | Off-Cloud Population | Galli et al. 2020 | 12 | 358.3 | -13.7 | 146 | 141 | 3 | -4 | 3 | -34 | 4 | 3.9 | 2.2 | -2.2 | 0.4 | 0.6 | 1.0 | 6 |
| Corona Australis | On-Cloud Population | Galli et al. 2020 | 6 | 359.9 | -16.7 | 149 | 143 | 3 | 0 | 1 | -43 | 2 | 4.6 | 2.1 | -1.4 | 0.4 | -1.0 | 0.6 | 5 |
| Serpens | Far South | Herczeg et al. 2019 | 3 | 27.3 | 2.4 | 384 | 341 | 3 | 176 | 0 | 16 | 0 | 12.3 | 2.6 | 1.9 | 1.7 | -3.3 | 0.2 | 3 |

**Extended Data Table 1: Summary of the 3D positions and 3D velocities of young stellar clusters within 400 pc of the Sun.** The rows are color-coded according to the cluster's longitude and map to the colors of the clusters shown in Figures 1 and 2. **(1)** Name of the cluster region. **(2)** Name of the cluster subgroup. **(3)** Original publication forming the basis for stellar membership. **(4)** Number of stellar members used to estimate the average cluster properties. **(5)** Mean longitude **(6)** Mean latitude **(7)** Mean distance **(8-9)** Mean x position and x uncertainty in Heliocentric Galactic Cartesian coordinates. **(10-11)** Mean y position and y uncertainty in Heliocentric Galactic Cartesian coordinates. **(12-13)** Mean z position and z uncertainty in Heliocentric Galactic Cartesian coordinates. **(14-15)** Mean U motion and U uncertainty along the Heliocentric Galactic Cartesian x direction in the LSR frame. **(16-17)** Mean V motion and V uncertainty along the Heliocentric Galactic Cartesian

y direction in the LSR frame. **(18-19)** Mean W motion and W uncertainty along the Heliocentric Galactic Cartesian z direction in the LSR frame. **(20)** Estimated age of the cluster drawn from the publication in column (3).

| Cluster | -16 UCL Born | -15 LCC Born | -14 SNe in UCL/LCC Make Bubble | -10 USCO and Oph "Old" Born | -6 CrA, Taurus "Old" Born | -5 CrA Born | -3 Taurus "Young", Lupus Born | -2 Taurus "Young", Cham Born | -1 Taurus "Young", Oph "Young" Born | Now Dense Gas Enveloping Bubble |
|---|---|---|---|---|---|---|---|---|---|---|
| UCL | (40,19,-24) | (44,14,-20) | (48,9,-17) | (64,-11,-3) | (81,-32,10) | (86,-37,14) | (95,-47,20) | (99,-51,23) | (104,-56,26) | (109,-61,29) |
| LCC | | (31,6,-18) | (32,0,-16) | (39,-27,-9) | (45,-54,0) | (47,-61,1) | (51,-74,5) | (53,-80,7) | (54,-87,9) | (57,-93,11) |
| Rho Oph "Old" (Population II) | | | | (75,-8,14) | (97,-11,28) | (102,-11,31) | (113,-12,37) | (119,-13,39) | (125,-13,41) | (131,-14,43) |
| Upper Scorpius (USCO) | | | | (83,-10,29) | (100,-14,42) | (105,-15,44) | (115,-17,49) | (119,-18,50) | (125,-18,52) | (130,-19,53) |
| Taurus (D3 - South) | | | | | (-92,-40,-22) | (-96,-32,-25) | (-103,-16,-30) | (-107,-8,-33) | (-111,-1,-35) | (-116,6,-37) |
| Corona Australis (Off-Cloud Population) | | | | | (120,9,-34) | (123,7,-34) | (130,2,-35) | (134,0,-35) | (137,-1,-35) | (141,-4,-34) |
| Taurus (C9 - 118TauEast) | | | | | (-89,13,-1) | (-92,9,-2) | (-98,2,-5) | (-101,0,-6) | (-104,-4,-8) | (-109,-7,-9) |
| Corona Australis (On-Cloud Population) | | | | | | (120,7,-34) | (129,4,-38) | (133,2,-40) | (138,1,-41) | (143,0,-43) |
| Lupus (4) | | | | | | | (129,-56,22) | (134,-59,22) | (139,-61,23) | (145,-63,23) |
| Taurus (C7 - L1527) | | | | | | | (-117,1,-25) | (-123,6,-28) | (-129,10,-30) | (-136,14,-33) |
| Taurus (C8 - B213) | | | | | | | (-127,16,-44) | (-135,19,-44) | (-143,21,-43) | (-151,24,-43) |
| Lupus (3) | | | | | | | (126,-48,24) | (132,-50,25) | (139,-52,25) | (146,-54,25) |
| Taurus (D2 - L1558) | | | | | | | (-108,-12,-39) | (-113,-8,-41) | (-118,-4,-43) | (-123,0,-45) |
| Lupus (Off-Cloud Population) | | | | | | | (128,-48,28) | (134,-49,29) | (140,-51,29) | (146,-52,29) |
| Taurus (D1 - L1544) | | | | | | | (-143,-8,-18) | (-152,-6,-19) | (-161,-5,-20) | (-170,-3,-21) |
| Taurus (C6 - L1524) | | | | | | | | (-110,6,-29) | (-116,9,-32) | (-123,13,-34) |
| Chamaeleon (1-South) | | | | | | | | (86,-152,-43) | (84,-156,-47) | (82,-161,-51) |
| Taurus (C5 - L1546) | | | | | | | | (-141,8,-46) | (-148,10,-46) | (-155,12,-45) |
| Taurus (D4 - North) | | | | | | | | (-118,6,-34) | (-123,12,-36) | (-128,17,-38) |
| Taurus (C1 - L1551) | | | | | | | | (-126,1,-50) | (-131,1,-49) | (-136,2,-49) |
| Chamaeleon (1-North) | | | | | | | | (85,-157,-40) | (84,-161,-44) | (84,-165,-49) |
| Chamaeleon (2) | | | | | | | | (107,-151,-46) | (106,-154,-47) | (105,-158,-48) |
| Taurus (C2 - L1495) | | | | | | | | | (-116,20,-32) | (-122,24,-35) |
| Rho Oph "Young" (Population I) | | | | | | | | | (126,-16,42) | (131,-15,40) |

**Extended Data Table 2: Temporal evolution of cluster births at the surface of the Local Bubble's expanding shell.** The time sequence from left to right shows the (x,y,z) positions (in pc) of clusters as a function of time. If a cell appears blank, the cluster has not yet been born, with the first filled cell (from left to right) indicating the birth position of the cluster. These birth positions are used to fit the analytic superbubble expansion model presented in the Methods section. All positions are given with respect to the Local Standard of Rest. The rows are color-coded according to the cluster's longitude and map to the colors of the clouds shown in Figures 1 and 2, as well as Extended Data Table 1. Significant events corresponding to each time step are summarized at top.

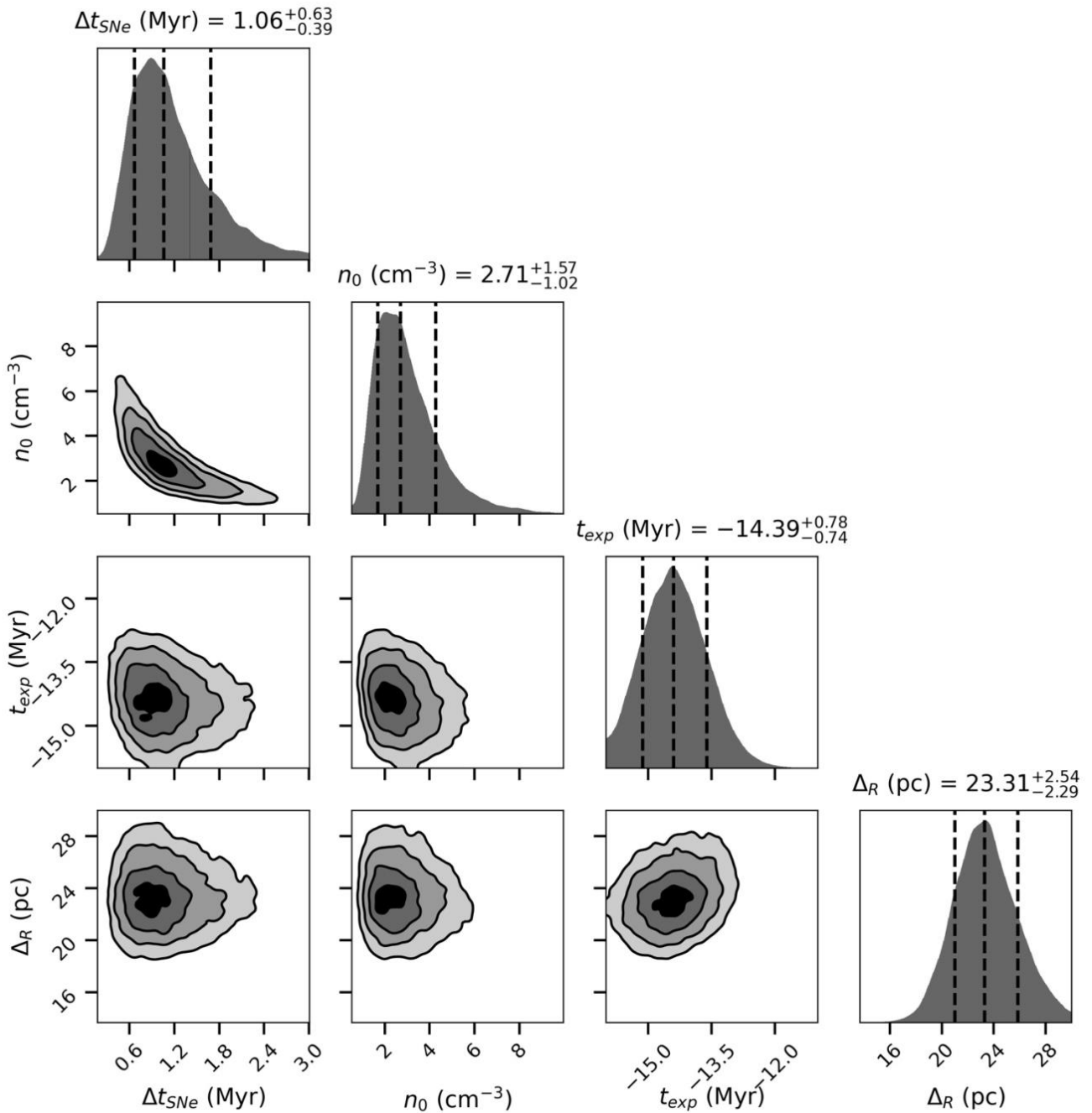

**Extended Data Figure 1: 1D and 2D marginal distributions ("corner plot") of the model parameters governing the evolution of the Local Bubble's expanding shell.** Parameters include the time since the first explosion (i.e. the age of the Local Bubble), $t_{exp}$, the density of the ambient medium the bubble is expanding into, $n_0$, the time between supernova explosions powering its growth, $\Delta t_{SNe}$, and the thickness/uncertainty on the expanding shell radius $\Delta_R$. In the 1D distributions, the vertical dashed lines denote the median and $1\sigma$ errors, while in the 2D distributions, we show the $0.5\sigma$, $1\sigma$, $1.5\sigma$, and $2\sigma$ contours.

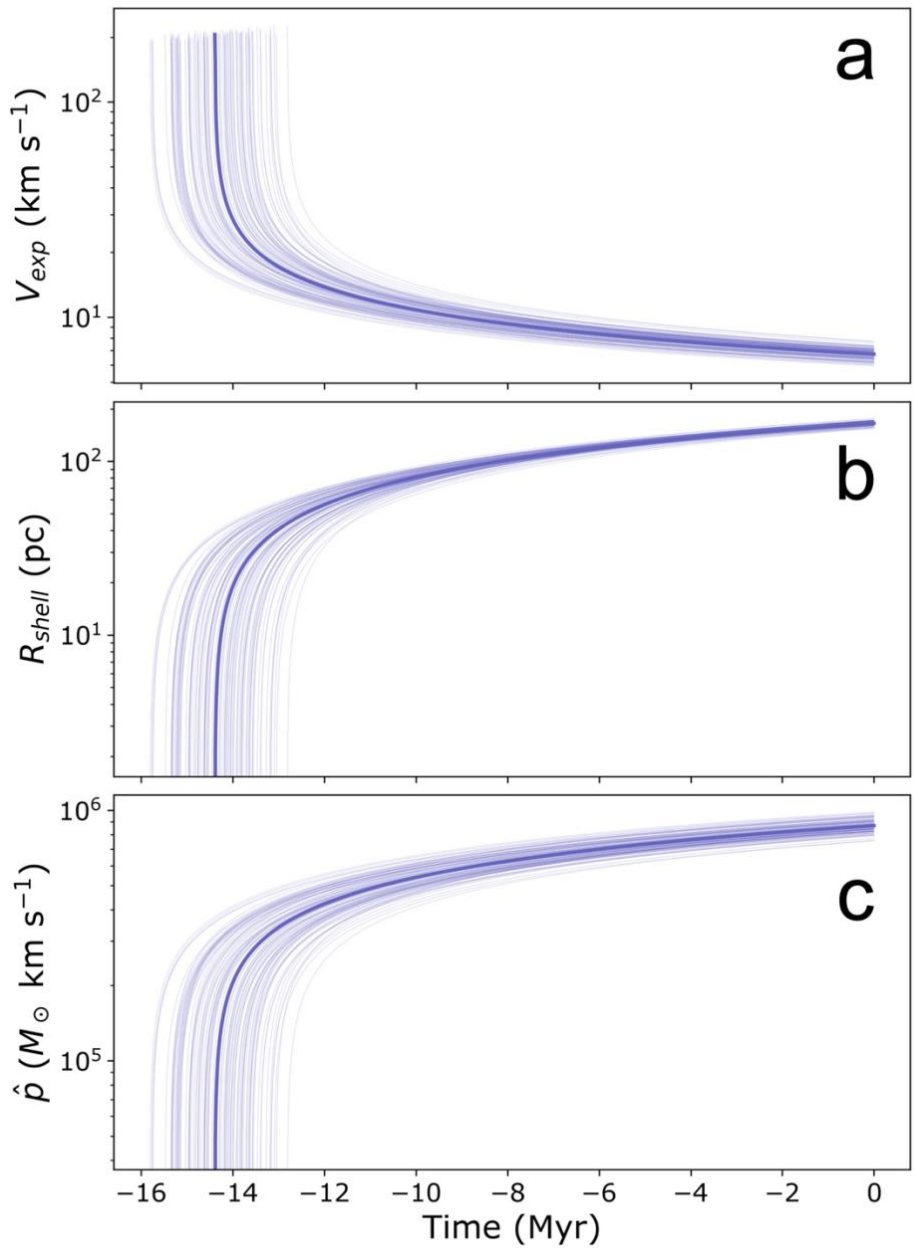

**Extended Data Figure 2**: **Temporal evolution of the Local Bubble, based on the fit to the dynamical tracebacks and the analytic expansion model[22] summarized in the Methods section. Panel a)** The evolution of the Local Bubble's expansion velocity $v_{exp}$. **Panel b)** The evolution of the Local Bubble's shell radius $R_{shell}$. **Panel c)** The evolution of the average momentum injection per supernova $\hat{p}$. The thick purple line represents the median fit, while the thin purple lines represent random samples. We estimate a current radius of $165 \pm 6$ pc and current expansion velocity of $6.7^{+0.5}_{-0.4}$ km/s, corresponding to time t=0 Myr (the present day).

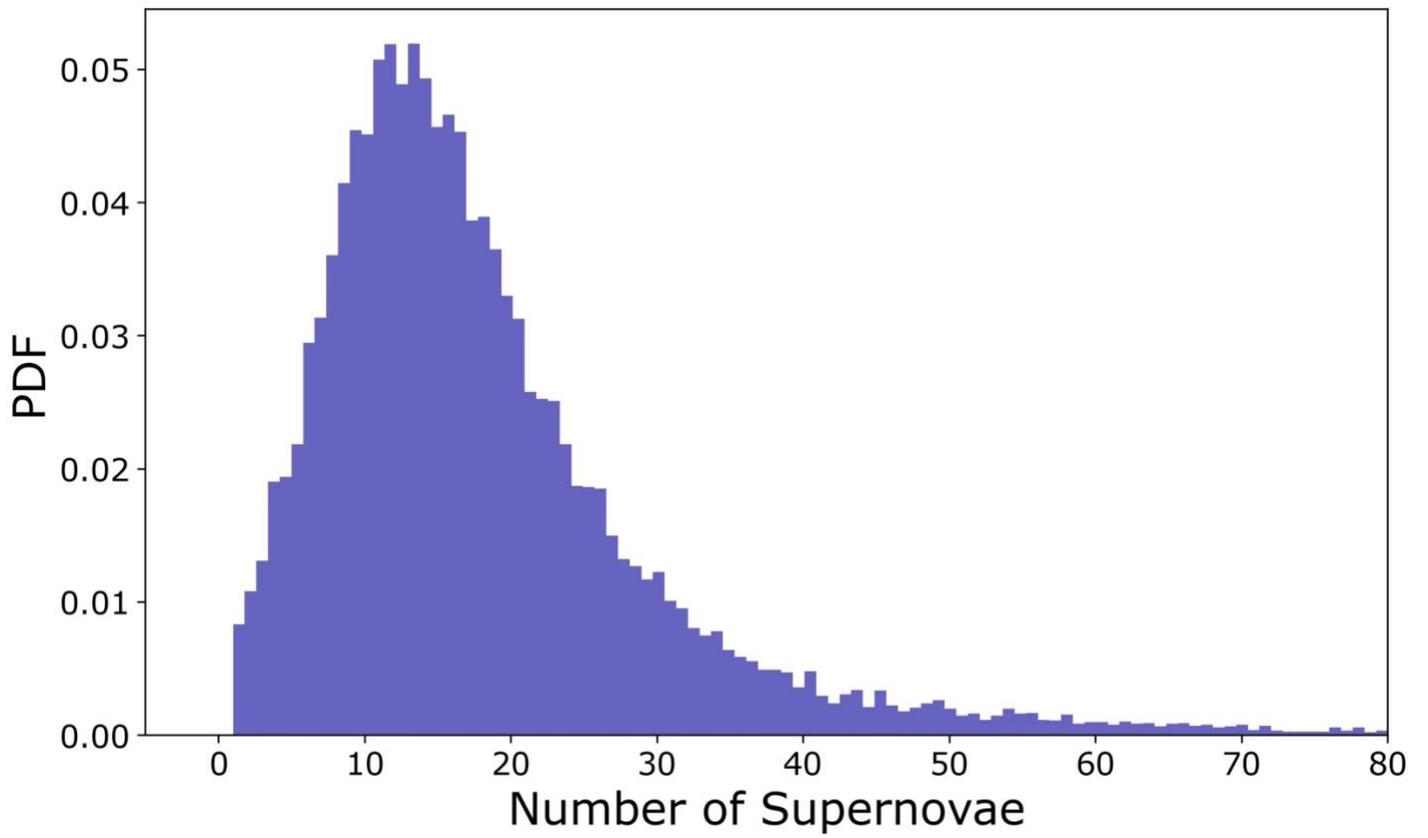

**Extended Data Figure 3**: **PDF of the estimate of the number of supernovae required to power the Local Bubble's expansion.** The estimate is obtained by comparing the shell's present-day momentum to the average momentum injected by supernovae.

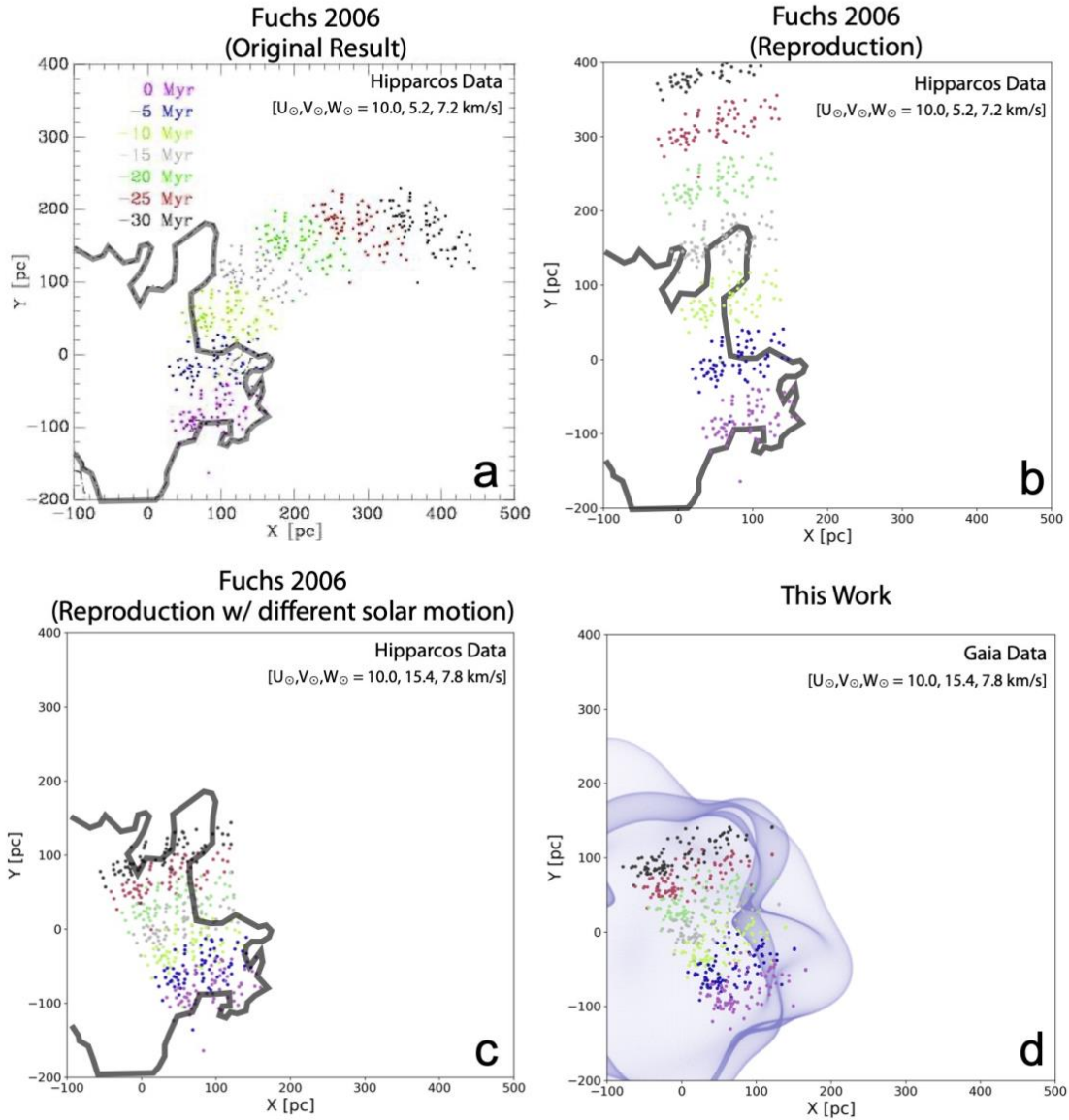

**Extended Data Figure 4**: **Analysis of the stellar tracebacks of the UCL and LCC clusters, whose progenitors were likely responsible for the supernovae that created the Local Bubble**. The scatter points indicate the positions of the current cluster members of UCL and LCC, which are colored as a function of time (spanning the present day in pink to 30 Myr ago in black). **Panel a**: Using Hipparcos data and adopting a solar peculiar motion $(U_\odot, V_\odot, W_\odot) = (10.0, 5.2, 7.2)$ km/s[46], prior literature[6,7] find that UCL and LCC were born outside the Local Bubble (black trace[4]) 15 Myr ago and only entered its present-day boundary in the past 5 Myr (reproduced from Figure 6, *The search for the origin of the Local Bubble redivivus*, Fuchs et al. 2006. *Mon. Not. R. Astron. Soc.* Volume 373, Issue 3, pp. 993-1003). **Panel b:** We attempt to reproduce the results from prior literature[6,7] using the same data and solar motion, but are unable to account for the curvature of the traceback, finding the UCL and LCC formed just inside its northern boundary 15 Myr ago. **Panel c:** Using a different value for the solar motion, $(U_\odot, V_\odot, W_\odot) =$ (10.0, 15.4, 7.8) km/s[41] but the same Hipparcos data, we find that UCL and LCC were born near the center of the Local Bubble. **Panel d:** Finally, using updated *Gaia* data but the same adopted solar motion used in panel c. $(U_\odot,$

V☉, W☉) = (10.0, 15.4, 7.8) km/s[41], we also find that UCL and LCC were born near the center of the bubble, given an updated model for its surface.[13]